\documentclass[journal,12pt,onecolumn,letterpaper]{IEEEtran}
\usepackage{arxiv}
\usepackage{geometry}
\usepackage{times}
\usepackage{cite}
\usepackage{url}
\usepackage{graphicx}
\graphicspath{{Images_SimplyMime/}}
\usepackage{lscape}
\usepackage{subfigure}
\usepackage{rotating}
\usepackage{rotfloat}
\usepackage{xcolor}
\usepackage{amsmath}
\usepackage{amssymb}
\usepackage[linesnumbered,ruled,vlined]{algorithm2e}
\usepackage{pseudocode}
\usepackage{array}
\usepackage[english]{babel}
\usepackage{gensymb}
\usepackage{textcomp}
\usepackage{placeins}
\usepackage{balance}
\usepackage{booktabs}

\title{Verifiable Passkey – The Decentralized Authentication Standard } 

\author{%
  
    Aditya Mitra \\
    CyberMACS,  \\
     Kadir Has University, Turkey \\
    \texttt{adityaarghya0@gmail.com}
\And
Sibi Chakkaravarthy Sethuraman\\
    Centre of Excellence, Artificial Intelligence \& Robotics (AIR),\\
    School of Computer Science and Engineering\\
    VIT-AP University, India \\
    \texttt{sb.sibi@gmail.com} \\

}

\begin{document}



\maketitle
\begin{abstract}
Passwordless authentication has revolutionized the way we authenticate across various websites and services. FIDO2 Passkeys, is one of the most-widely adopted standards of passwordless authentication that promises phishing-resistance. However, like any other authentication system, passkeys require the user details to be saved on a centralized server, also known as Relying Party (RP) Server. This has led users to create a new passkey for every new online account. While this just works for a limited number of online accounts, the limited storage space of secure storage modules like TPM or a physical security key limits the number of passkeys a user can have. For example, Yubico Yubikey 5 (firmware 5.0-5.6) offers to store only 25 passkeys, while firmware 5.7+ allows to store upto 100. [1]. To overcome this problem, one of the widely adopted approaches is to use Federated Authentication with Single Sign On (SSO). This allows the user to create a passkey for the Identity Provider (IdP) and use the IdP to authenticate to all service providers. This proves to be a significant privacy risk since the IdP can potentially track users across different services. To overcome these limitations, this paper introduces a novel standard ‘Verifiable Passkey’ that allows the user to use Passkeys created for a Verifiable Credential issuer across any platform without risking privacy or user tracking. 
\end{abstract}




\enlargethispage{10pt}

\section{Introduction}

It was in the 1960s that Fernando Corbato developed the first password authentication system for MIT Compatible Time-Sharing System (CTSS). As the threats evolved, so did the authentication systems, giving way to 3 commonly used factors of authentication: Knowledge based factors (memorized secrets), Possession based factors and Inherence based factors. NIST defines authentication as “Verifying the identity of a user, process, or device often as a prerequisite to allowing access to resources in an information system \cite{nist_fips_2006}". 
Due to the rise of social engineering-based attacks like phishing, it has been quite necessary to use a phishing-resistant method of authentication. FIDO2 Passkeys, jointly developed by FIDO Alliance and W3C is a promising standard for phishing resistant authentication and has been widely adopted by various web services and is supported by most modern operating systems and browsers. ‘Windows Hello’, offered by Microsoft, ‘iCloud Keychain’ offered by Apple and ‘Google Password Manager’ offered by Google are leading examples of supported platforms. FIDO2 uses possession-based factors for authentication where the claimant needs to cryptographically prove the possession of a device like a physical security key, a specific device like a smartphone or a PC containing a cryptographic element, etc. to be authenticated. FIDO2 assures phishing resistance by binding each cryptographic secret to the RP Server ID (a registrable domain suffix of or equal to the origin’s effective domain) \cite{webauthn_l2_2021} where the passkey is created. This ensures no malicious domain name, usually used for phishing can bind to the passkey created for the original domain.
This, while ensuring phishing resistance, also implies the user needs to create a new passkey for every online service he uses. For a limited number of accounts, it works just fine but when the number of passkeys per user increases this may lead to a problem due to the limited memory of the secure storage. For example, YubiKey 5 (a market leading physical security key by Yubico), for firmware versions 5.0 to 5.6 supports holding only 25 passkeys while firmware 5.7 and up supports up to 100 only \cite{yubico_support_2025}. 
To overcome this problem, one of the well adopted measures is to use Federated authentication where the user needs to enroll passkey for only the Identity Provider (IdP) server and other service providers could use Single Sign On, OAuth/ OIDC or any other Federated Authentication standard. This introduces another problem that IdP can track the user across the various services he uses, and this introduces a privacy risk. 
The proposed standard proposes the use of Verified Credentials to share Passkey credentials to service providers who can then verify the passkey credentials and use the same to authenticate the user. 

 We make the following contributions:
\begin{itemize}
    \item A novel way to use passkeys with verifiable credentials
    \item 
   Verifiers would not create new passkeys, hence not use up the storage of secure media 
    \item Fully compliant with present browsers and platforms
\end{itemize}

\section{Literature Review and Research Gap}
FIDO2 has emerged as a standard for passwordless authentication promising phishing resistance. It has been adopted by various websites and services, eliminating the needs for passwords and knowledge-based authentication systems \cite{webauthn_l2_2021}. It has provided cryptographic authentication standards using platform and roaming authenticators, thus making it easy to use and adopt \cite{fido_password_trends_2025, ramat_passkey_experience_2025 }.  However, a passkey created for one website may not be used on another, due to strong Relying party requirements of the standards. This led users to have to sign up for multiple user accounts for various websites. The other alternative was to use federated authentication techniques like Single Sign Ons.
However, federated authentication systems meant a form of vendor lock-in. The authentication provider could control access to where the user can authenticate to. Further, it is quite possible for a federated authentication provider to log user authentication and analyze the same in a form of surveillance. Further, even personally identifiable information (PII) can be leaked by the federated authentication providers \cite{wursching_fido2_2023, jannett_sso_monitor_2024}.
Verifiable Credentials provide a promising approach to decentralization of information where the user owns the credentials, yet it is signed by an issuer and can be verified by a verifier independently. This reduces the reliance on any centralized infrastructure for the information to be deemed valid \cite{westers_sso_privacy_2024, pham_pii_leakage_2025}. However, verifiable credentials were not primarily built for authentication, it was rather built for verifiability of documents without a centralized infrastructure. The proposed standard aims to bridge the gap between Verifiable Credentials and Passkeys to develop an authentication standard where the user holds the credentials, which can be used across various websites and platforms but may not be tracked by any centralized party.
Verifiable Passkeys aim to be compliant with the NIST SP-800-63 authentication standards \cite{vc_overview_2025, vc_data_model_v2_2025}. It highlights the usability and compliance, providing over AAL2 level of authenticator assurance due to the use of FIDO2 standards and strong authentication. It at the same time minimizes the risk of centralized data logging and metadata-based surveillance.

\section{Advantages of Verifiable Passkeys}
Verifiable Passkeys is a seamless Verifiable Credentials based passwordless authentication standard that allows the user to supply his own credential to the verifier. This is followed by phishing resistant authentication to ensure the user is who he claims to be. This uses cryptographic proofs as per the WebAuthn \cite{webauthn_l2_2021}. It reduces the dependence on server-side credential storage, and it makes authentication decentralized yet strong. 
The user is supposed to store his own verifiable credential and present the same during authentication to any service provider or verifier. This also ensures that the credentials would not be leaked even in case of a breach on the server side. Table. \ref{Tab:vp_comparison} highlights the differences of Verifiable Passkeys with other commonly used authentication standards.

\begin{table}[ht]
\centering
\caption{Comparison of Verifiable Passkey with other standards}
\label{Tab:vp_comparison}
\begin{tabular}{p{4cm} p{2cm} p{5cm} p{2.75cm} p{2cm}}
\hline
\textbf{Factor} & \textbf{Passwords} & \textbf{Passkeys} & \textbf{Federated Authentication (with passkeys)} & \textbf{Verifiable Passkey} \\
\hline
Phishing resistance & No & Yes & Yes & Yes \\ \\
Storage constraints limit usability & No & Yes & No & No \\ \\
Privacy risk of user being tracked across services & No & Partial (Yes if passkeys are stored in centralized passkey managers, no if hardware-backed) & Yes & No \\ \\
Server-side storage required & Yes & Yes & Yes & No \\ \\
Risk of server-side breach & Yes & Yes & Yes & No \\
\hline
\end{tabular}
\end{table}

\section{Technical details of Verifiable Passkey}
\subsection{Primitives}
\begin{itemize}
\item User: It is the entity that is trying to authenticate to the service provider. The user is also known as ‘claimant’.
\item Service provider: It is the app or web service where the user is trying to authenticate to. It is also known as ‘verifier’.
\item Device: It is a cryptographic device that is in possession of the user. This device can be a standard smartphone, PC or physical security keys. This device usually communicates with the user’s browser via Client to Authenticator Protocol (CTAP) as defined by FIDO Alliance. It is also known as ‘authenticator’.
 \item Browser: This is a standard web browser that supports W3C WebAuthn and CTAP protocols. WebAuthn is used for the authentication on browser while CTAP is used to communicate with the cryptographic device.
  \item Issuer: It is the entity that enrolls the user and generates a verifiable credential from the passkey.
\item Challenge: This is a cryptographic nonce of 16 bytes or more. It is used for authentication in challenge-response methodology.
\item PageX: This is a special webpage hosted by the issuer. It is a static webpage and does not call any api on the issuer server to prevent tracking. It takes in a cryptographic nonce from URL parameters and performs a ‘get assertion’ call solely on the client side. This returns an assertion value from the cryptographic device and it calls a redirect URI with this assertion in URL parameters. It is essential that it does not call any API of the issuer and works on only client side. The purpose of this is to work with the Relying Party ID constraint that the passkey will accessible only from a page hosted by the issuer, yet the passkey is to be used by the verifier. Ideally the PageX is to be hosted by non-profit organizations and be open source and auditable.

\end{itemize}

The user, in order to use the Verifiable Passkeys, user must need to enroll with the issuer first. The algorithm for the same might be represented as:

\begin{itemize}
\item User opens Issuer webpage and enters details
\item Issuer server generates Public Key Credential Creation Options with a challenge and redirects to PageX with the same.
\item 	PageX uses CTAP to send a ‘make credential’ command to the cryptographic device with the challenge.
\item	The cryptographic device generates a new keypair.
\item	The cryptographic device signs the challenge in the options.
\item	The cryptographic device returns the signed challenge and the public key to the browser in the form of an attestation via CTAP.
\item	The PageX returns attestation to the issuer.
\item	The issuer validates the signed challenge with the public key. 
\item	The issuer generates a verifiable credential containing the user details and the public key, signed with the private key of the issuer.
\item	The issuer sends the verifiable credential to the browser.
\item The user downloads the verifiable passkey and keeps it for later usage.

\end{itemize}

Figure 1 shows the workflow for passkey enrollment and verifiable credential generation.

 \begin{figure}[htbp]
\centering
\includegraphics[width=1\textwidth]{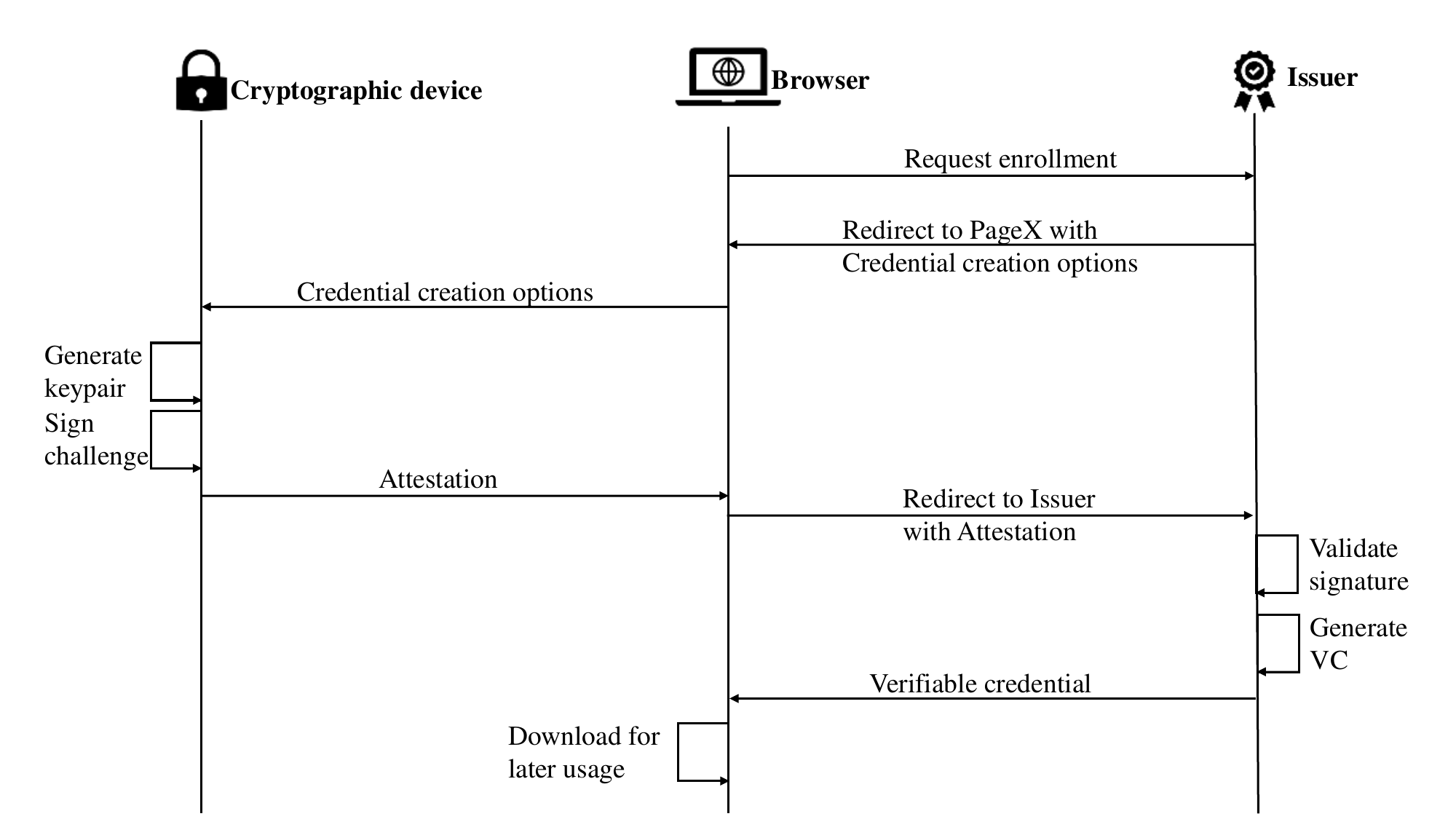}
\caption{Enrollment workflow}
\label{FIG:Enrollment_Workflow}
\end{figure}

Once the verifiable credential is created, the user can save it locally or in any wallet. The user may then present it to verifier, followed by performing passwordless authentication to prove possession of the cryptographic device, hence authenticating. 

The workflow of authentication is as:

\begin{itemize}

\item	User opens Verifier login page
\item	User uploads his VC
\item	Verifier verifies the validity of the VC
\item	Verifier extracts PageX URL from the VC
\item	Verifier creates a challenge
\item	Verifier redirects to PageX with the challenge
\item	PageX creates Public Key Credential Request Options on client side
\item	PageX invokes the cryptographic device and to get assertion
\item	PageX redirects back to verifier redirect URI with the assertion
\item	Verifier extracts the attested credential public key from the VC
\item	Verifier validates the assertion against the attested credential public key to authenticate the user

\end{itemize}

Figure 2 shows the authentication flow of a user.

\begin{figure}[htbp]
\centering
\includegraphics[width=1\textwidth]{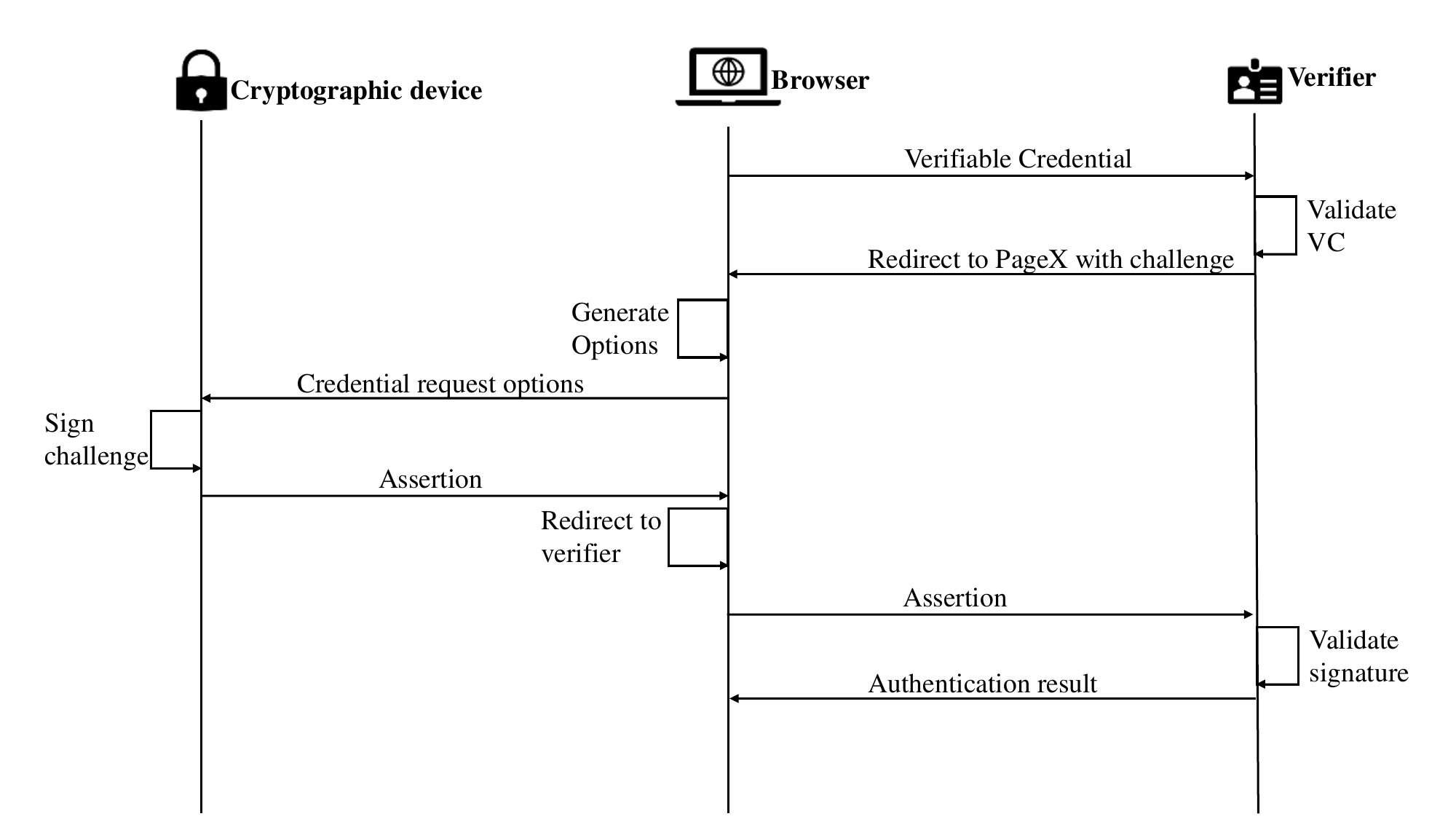}
\caption{Authentication Workflow}
\label{FIG:Authentication_Workflow}
\end{figure}

Hence, it is evident that the issuer server cannot track the authentications, unlike federated authentication, which is a major positive move towards protecting privacy of the user.

The structure of a verifiable passkey is as follows:
\begin{verbatim}
{
  "@context": [
    "https://www.w3.org/2018/credentials/v1"
  ],
  "id": "http://example.org/credentials/3732",
  "type": [
    "VerifiableCredential",
    "VerifiablePasskey"
  ],
  "issuer": "did:web:AdityaMitra5102.github.io/VPass-Issuer",
  "issuanceDate": "2025-11-24T18:19:08.216905",
  "credentialSubject": {
    "id": "User creds",
    "user": {
      "email": "adityaarghya0@gmail.com",
      "name": "Aditya Mitra",
      "phone": "+91xxxxxxxxxx"
    },
    "pagex": "https://adityamitra5102.github.io/VerifiablePasskey/pagex.html",
    "cred": {
      "aaguid": "nd0YF69aRnKiuT492VAAqQ==",
      "credential_id": "R+NuXRSywj4RPGGUpR+cuap7YIs2WCBnItvNZgS+4yM=",
      "public_key": {
        "1": 2,
        "3": -7,
        "-1": 1,
        "-2": "base64_zHFxjAiAduj7MrqDQBIIjh/99t42khQt0IUchij5xCE=",
        "-3": "base64_sx9hzYecDFcgyMZ1fcu1obA4oc3rN9KjgNz1m5I7MVA="
      }
    }
  }
}
\end{verbatim}

As evident from the sample Verifiable Passkey, the Passkey credential is serialized and added to the Verifiable Passkey so that it can be used by other verifiers. It is achieved by encoding the AAGUID and Credential ID to Base64 string, while the public key is dumped to COSE type. The parameters which are of type integer, or number are kept as is, while the parameters of type bytes are encoded to Base64 with the magic word 'base64\_' prepended. Similar deserialization method is used on the verifier side to make the passkey usable for authentication. 

After creating the Verifiable Credential, the issuer creates a Verifiable Presentation of the same, which the user can use with Verifiers. Further, it is to be noted that every assertion and attestation values contain a ‘clientDataJSON’ field as per the Web Authentication specifications. It mentions the domain of the page which triggers the authentication, PageX in this case, which is a part of the cryptographic signature. This can be verified by both the issuer and verifier to ensure the action was performed by a legitimate PageX. It is optional to be verified for the Verifier because once the Issuer verifies it, it is already cryptographically bound to the FIDO2 credential as per the WebAuthn specifications.

\section{Implementation}
The proposed standard has been implemented and tested locally. Open-source python-based issuers and verifiers were created for the test. PageX was a static HTML page hosted on Github Pages. The issuer and verifier ran on independent machines, on localhost, to confirm they need not communicate with each other. The public key of the issuer was hosted on a separate Github repository which made it accessible to the verifier. 
For the first test, the passkey was created on platform authenticator on Windows Hello and the verifier was run on the same machine. For another test, the passkey was created on a roaming authenticator, on a physical security key and the verifier was tested on a different machine.
Both tests worked as expected and the Figures 3 to 9 highlight them. Figure 3 shows a user attempting to create a Verifiable Passkey from an issuer. Figure 4 shows PageX being used for credential creation. Figure 5 shows the system prompt from requesting to use a Hardware backed physical security key, though platform authenticator is supported as well. Figure 6 highlights the success message prompting the user to download the verifiable passkey. Figure 7 shows the contents of the verifiable passkey. Figure 8 shows the verifier verifying the passkey to retrieve user details, and figure 9 shows a successful authentication with the Verifiable passkey.

\begin{figure}[htbp]
\centering
\includegraphics[width=0.5\textwidth]{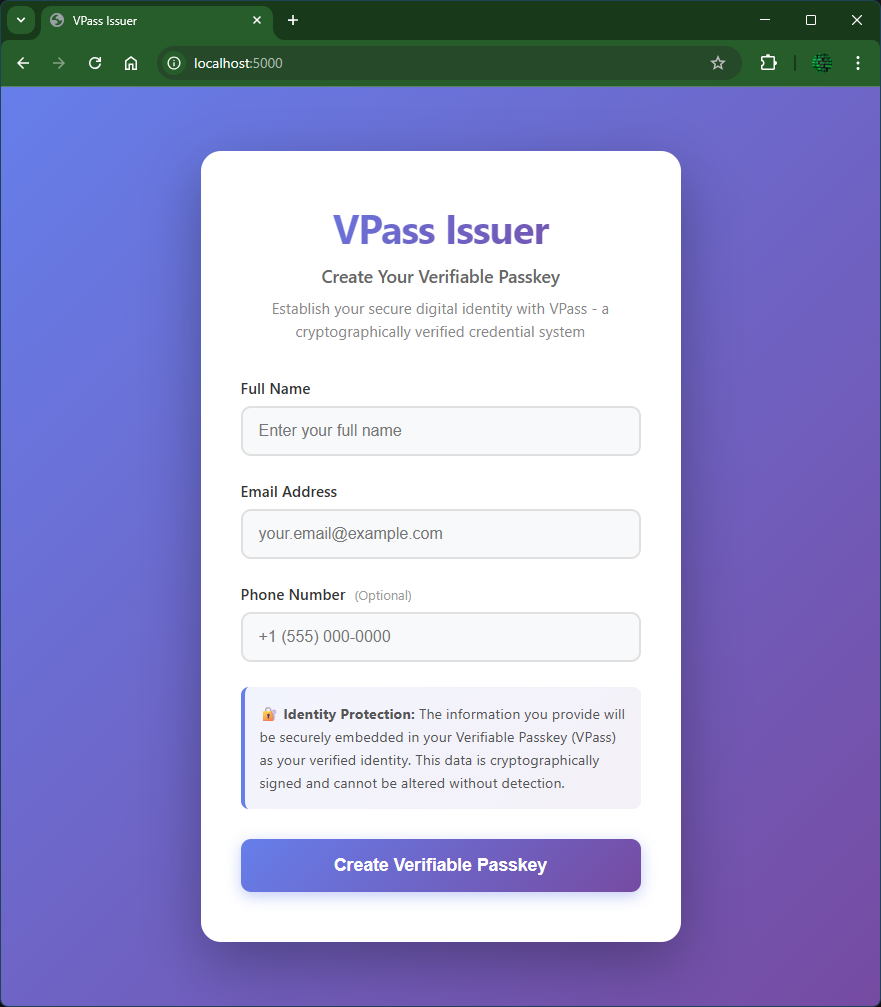}
\caption{Verifiable Passkey Issuer screen}
\label{FIG:Verifiable_Passkey_Issuer_screen}
\end{figure} 

\begin{figure}[htbp]
\centering
\includegraphics[width=0.5\textwidth]{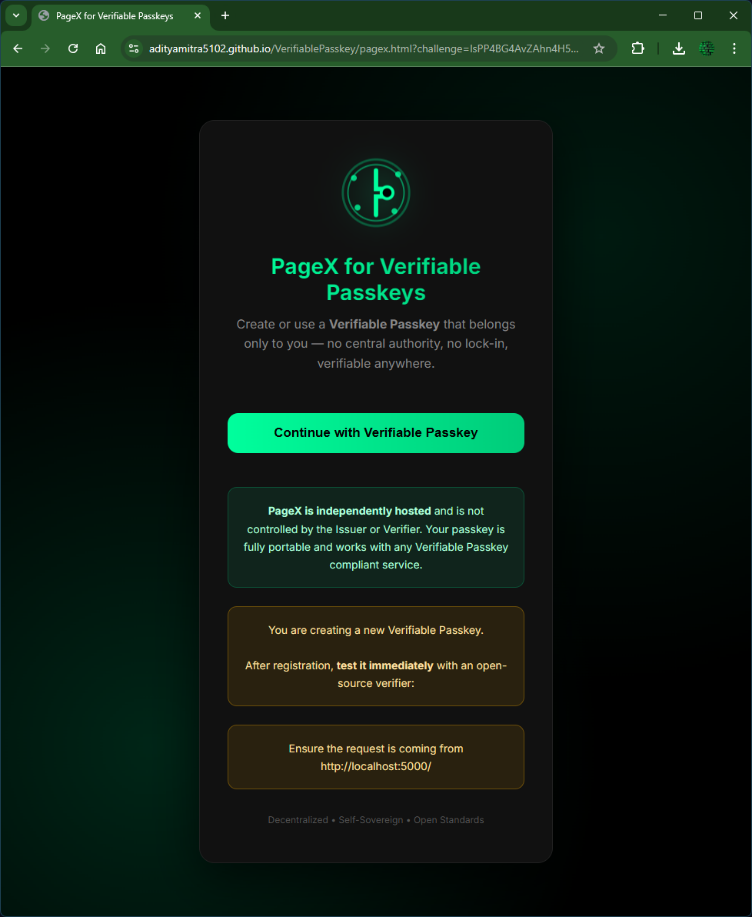}
\caption{PageX for creating Verifiable Passkey}
\label{FIG:Pagex_Verifiable_Passkey}
\end{figure} 

\begin{figure}[htbp]
\centering
\includegraphics[width=0.4\textwidth]{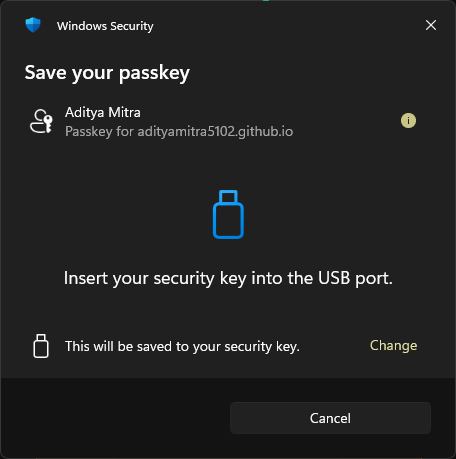}
\caption{System prompt for secure storage usage}
\label{FIG:System_Prompt}
\end{figure} 

\begin{figure}[htbp]
\centering
\includegraphics[width=0.5\textwidth]{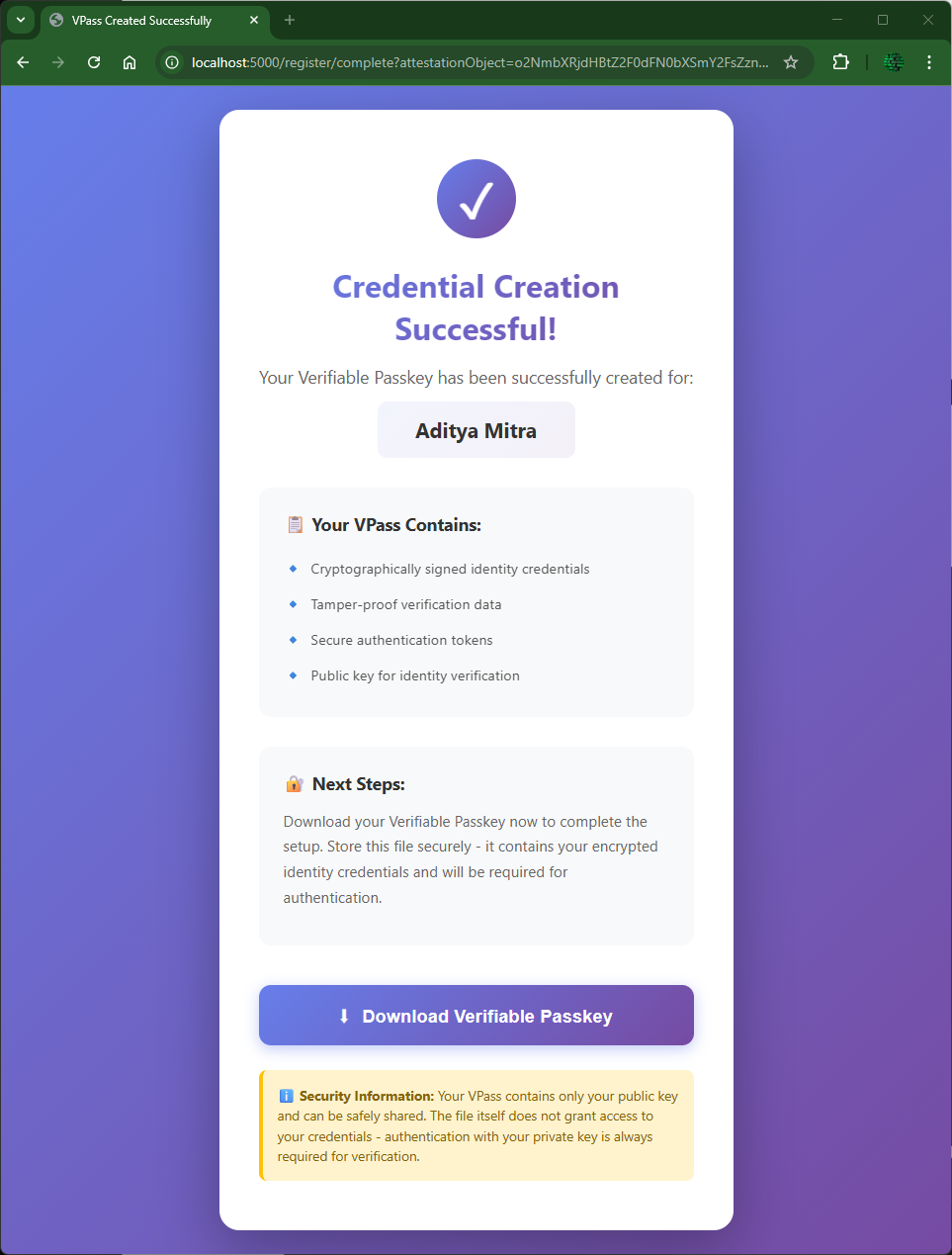}
\caption{Verifiable Passkey creation success}
\label{FIG:Verifiable_Passkey_Creation_Success}
\end{figure} 

\begin{figure}[htbp]
\centering
\includegraphics[width=0.5\textwidth]{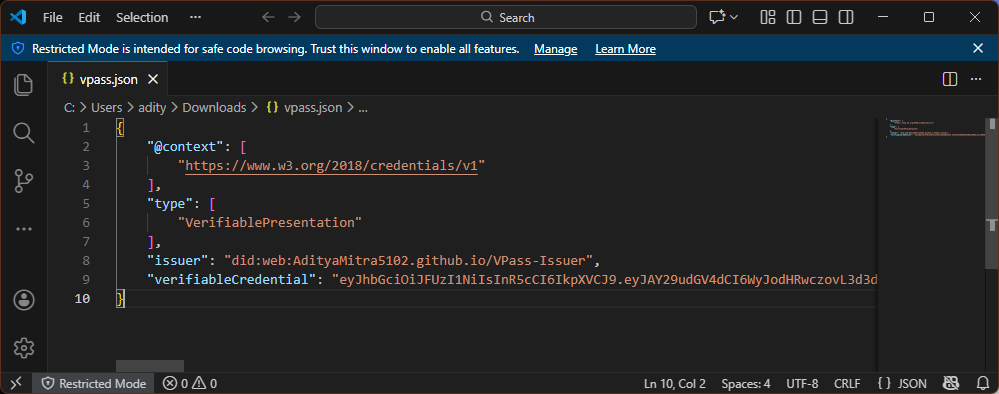}
\caption{Verifiable Passkey content}
\label{FIG:Verifiable_Passkey_Content}
\end{figure} 

\begin{figure}[htbp]
\centering
\includegraphics[width=0.5\textwidth]{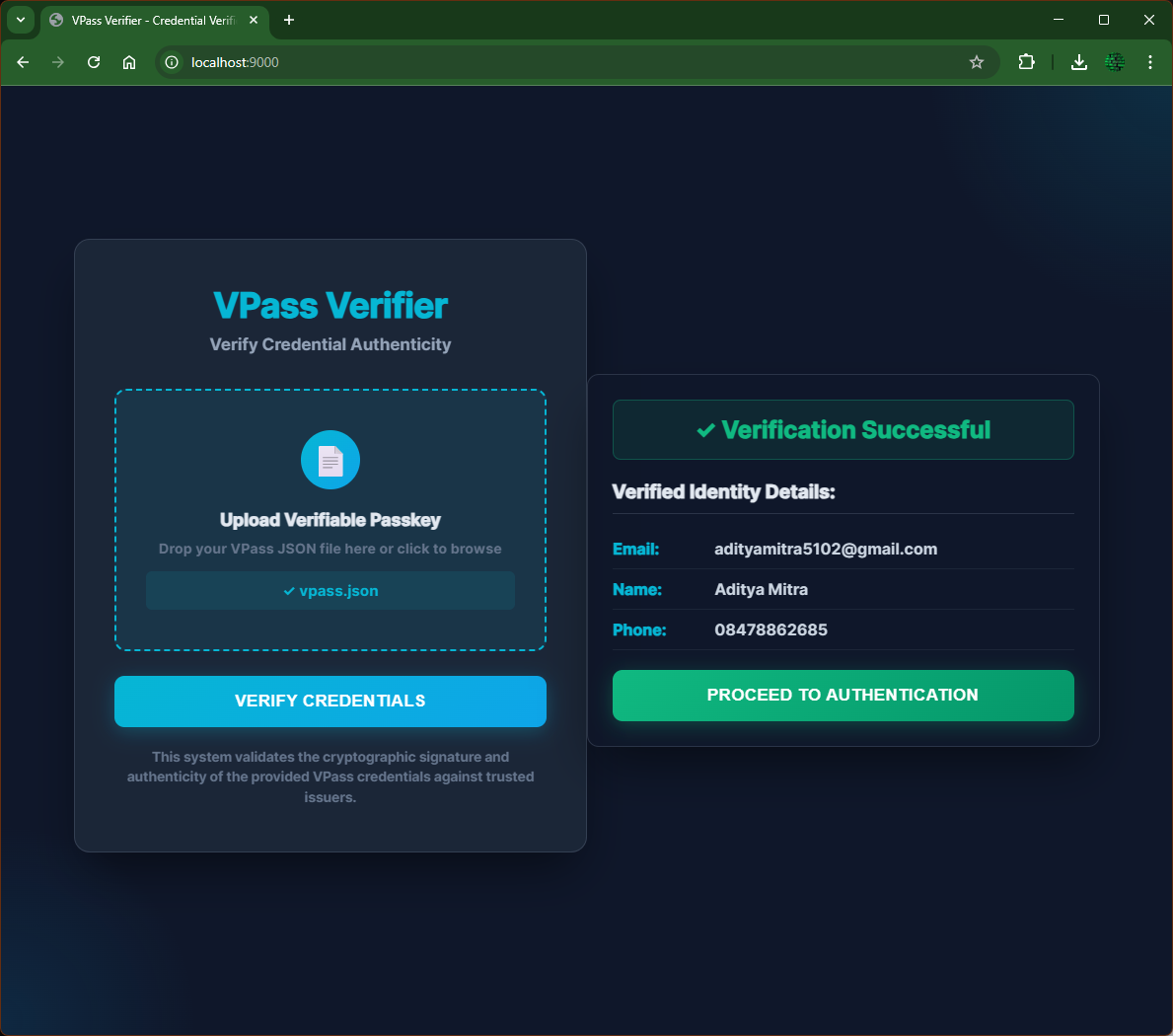}
\caption{Verifier verifying the Verifiable Passkey}
\label{FIG:Verifier_Verifiable_Passkey}
\end{figure} 

\begin{figure}[htbp]
\centering
\includegraphics[width=0.5\textwidth]{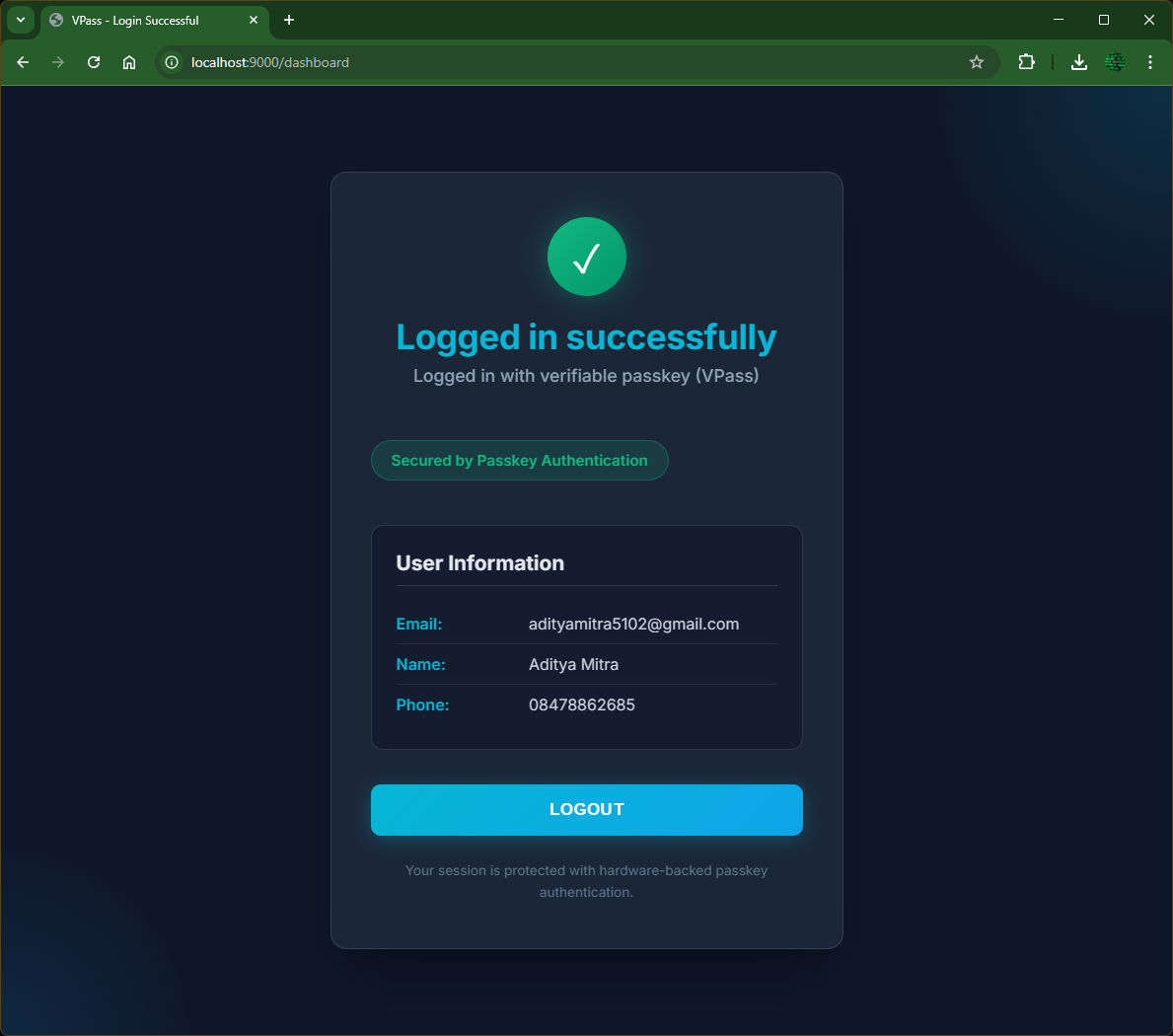}
\caption{Authentication successful with Verifiable Passkey}
\label{FIG:Authentication-Success_Verifiable_Passkey}
\end{figure} 

\section{Security Analysis and Threat Model}
Verifiable Passkeys have a robust threat model that extends the one of traditional passkeys, making it more usable and freer from vendor lock-ins. The threat model is highlighted as follows:

\begin{itemize}
\item	Phishing resistance: Verifiable Passkeys use passwordless authentication with the FIDO2 standards, which removes the reliance on memorized secrets and hence is phishing resistant.
\item	Credential stuffing and brute force: Verifiable Passkeys use cryptographic authentication techniques which mitigates any attempt to brute force.
\item	Reliance on SSO and single point of failure: Users relying on SSOs for passwordless authentication across multiple websites and services often have a single point of failure of the SSO service. Verifiable Passkeys make the users the owner of their credentials and there is no single point of failure.
\item	Surveillance and tracking: SSO and other federated authentication mechanisms often log where the credential is used, and the metadata can be extensively used for surveillance. Verifiable Passkeys remove the need of such centralized authorities and mitigate surveillance.
\item	Zero Trust: The verifier validates the verifiable passkey with the public key of the issuer before allowing the user to use it. This mitigates any risk of using malicious or fake credentials to authenticate a user.
\item	Zero knowledge: The verifier may not need to have all information about the user, other than the public key to authenticate the user. This can have various use cases in the modern privacy landscape like digital identities.
\item Assertion and Attestation details leaked through browser history: The assertion and attestation information does not contain anything sensitive. It contains only public keys (in case of assertion) and signed challenge. Hence, those getting exposed are not real threats.
\item Compromised PageX: The Issuer can verify the identity of the PageX used cryptographically from the attestation and ensure it is the intended PageX.
\item Credential theft: Even if an attacker is able to get the Verifiable Passkey file, the attacker cannot authenticate. 
\item User binding: The verifier cryptographically establishes trust with the issuer over decentralized identity protocols and to verify the validity of the Verifiable Passkey and then binds it to the user.
\item Replay attacks: A malicious attacker in possession with the verifiable passkey file of a legitimate user would still not have access to the cryptographic secrets or private key of the user. Hence, he would not be able to use the Verifiable passkey.
\end{itemize}

For example, a use case for Verifiable Passkeys can be in digital identities and age verification, which is of high priority in various countries at the time of writing this paper. The government may be able to issue a verifiable passkey with proof of age, signed by the government root of trust, while other websites or social media platforms may use this verifiable passkey as an additional factor of authentication to ensure the age of the user is verified. This allows the governments to create a secure, reusable, verifiable record for the citizens while the online platforms would be able to verify the same without having to take sensitive data like government IDs or driving licenses from the users to verify the age. This preserves the privacy of the user while ensuring compliance with government standards and protects the user from further data breaches.

\section{Conclusion and Future Scope}
This paper presents a standard that allows the export of passkeys as Verifiable Credentials and is used in various verifiers while maintaining the privacy of user. This prevents user tracking across various services. The proposed standard is considered to be more privacy preserving and secure as compared to federated authentication. It gives the user full control over his credentials and which one he wants to present at which service. FIDO2 based authentication takes the best of passwordless and provides phishing resistance while this standard maintains the decentralized nature of authentication. This is truly verifier-agonistic since any application can use Verifiable Passkeys to authenticate a user while ensuring the user can use the same VC across services.

\section{Acknowledgements}
The authors would like to thank Dr. ArulMozhi Varman, Vice Chancellor, VIT-AP University, and  Jagadish Chandra Mudiganti, Registrar, VIT-AP University. Special thanks to the team members of the Centre of Excellence, Artificial and Robotics (AIR), VIT-AP University. The source code of this research is available at \cite{VerifiablePasskey, VPassIssuer, VPassVerifier}.

\bibliographystyle{elsarticle-num} 
\bibliography{VP-arxiv}


\end{document}